# APPLIED STATISTICS: A REVIEW

By D. R. Cox

*Nuffield College, Oxford*

The main phases of applied statistical work are discussed in general terms. The account starts with the clarification of objectives and proceeds through study design, measurement and analysis to interpretation. An attempt is made to extract some general notions.

**1. Introduction.** The extremely welcome founding of the *Annals of Applied Statistics* prompts the question: what are the central principles of applied statistics? The great variety of the fields in which statistical considerations play some part makes it hazardous to attempt an answer. Any simple recommendation along the lines *in applications one should do so and so* is virtually bound to be wrong in some or, indeed, possibly many contexts. On the other hand, descent into a yawning abyss of vacuous generalities is all too possible.

The following comments are relevant mostly to applications in the natural and social sciences and the associated technologies, including medicine and health, although not confined to particular fields. I am broadly very sympathetic to the idea that success often hinges more on scientific sense than on technical mastery of complex methods; see, for example, Chatfield (1995). The literature of our field is, however, inevitably dominated by detailed issues of methodology. Some of the points that follow are expressed in rather different form by Cox and Snell (1980), Part I.

One general issue is the desirability of an intimate union between subject-matter and statistical aspects of an investigation. This theme runs implicitly throughout much of the following discussion. In the formulation of objectives and in discussions of design the need is quite clear. In analysis of data there can be tensions. At one extreme, in subjects such as those parts of physics with a strong theory base, it would be foolish to use statistical models not directly consistent with that base. On the other hand, in other fields there may be firmly held views, enshrined in tradition, that are both inconsistent with the particular data under analysis and in actual fact not well established by evidence. Another possibility is that the pioneering subject-matter paper in a field may have used a statistical method which, while adequate at the time, is not good for general use. There may be considerable pressure within the field to use the inferior method. In the former case, there is a clash between superficial subject-matter concerns and the ultimately empirical aethos of statistics and indeed, of science, which by and large respects the primacy of empirical information. In the latter case persistent comparison of older and newer methods may produce change.

In those parts of statistical analysis based on explicit probability models, the principle of sufficiency in a sense helps release some of the tension. Sufficiency divides the data into two components. The sufficient statistic itself summarizes information assuming the model to be essentially correct, whereas the rest of the data, taken conditionally on the sufficient statistic, allows model criticism and development. There are exceptions where the model is saturated with parameters, as for example in some forms of two-level factorial experiment. Then no information is available for model development; such models are occasionally very fruitful but their use is manifestly hazardous.

In analyses that are purely descriptive or are based on algorithmically specified methods [Breiman (2001)], the broad principle should be the same; the data typically provide some information about the appropriateness of the description or algorithm and this aspect should be examined, even if only informally. It is interesting and perhaps surprising that J. W. Tukey, who had an extraordinarily wide-ranging knowledge of the natural sciences down to fine detail, favored largely ignoring that knowledge in the main phases of analysis, introducing it only in the final stages of interpretation [Tukey (1975 approx), personal communication].

**2. Formulation of objectives.** While settling the objectives of an analysis is initially a subject-matter concern, sometimes an important and indeed crucial role of statistics is in clarifying what reasonably can and cannot be investigated with a given body of data, data either already available or projected. Alternatively, what extraneous assumptions will be needed in addition to the data in order to address some specific issue? Efron [(1998), Section 4.4], has remarked that while frequentist formulations of analysis are typically best suited for careful assessment of the strength of evidence, Bayesian formulations may allow the insertion of additional information which may open the route to bolder speculation, sometimes providing an alternative to sensitivity analysis.

A conventional division of objectives is between decision-making and inference, the latter in the sense of gaining understanding and assessment of evidence. While many investigations, including, in particular, most that one would regard as technological in some very broad sense, have a decision-making ultimate objective, this is by no means the same as requiring an automatic decision-rule that will dictate the decision once the data are available. Major decisions typically require review of the evidence and discussion of its strengths and weaknesses, much as in a formal inference problem. From this perspective the theoretical representation of all statistical methods as being concerned with decision-making is seriously flawed.

Problems of decision-making in the narrower sense are, in principle, most satisfactorily treated, where possible, in a fully Bayesian setting, that is, by inverse probability with an evidence-based prior and a specified utility function. There are, however, a number of strategical issues connected with applications. Suppose that the decision rule is used repeatedly. How is its performance to be checked and possibly updated and improved, formally possible within a limited setting by some version of Kalman filtering? Also, can the whole formalization be checked in an analogue of model criticism? At the formalization stage is it better to go straight for good empirical performance or is this better preceded by an attempt at a deeper analysis?

A key issue for decision procedures intended for broad application is stability under systematic changes of environment of application and over time. Even relatively complicated methods of cross-validation rarely address this satisfactorily. The concern with meaningful stability underlies also some of the issues of analysis outlined below.

Where the objective is not immediate decision-making but the improvement of understanding, research questions of various levels of ambitiousness are possible. At one level a fruitful objective may be the concise description of patterns of variability, perhaps by a formal model, perhaps by a series of graphs, tables and summarizing statistics. Interpretation is then left to informal arguments. Some applications of spectral analysis are of this kind. A more specific example concerns the description of spatial-temporal rainfall fields. Meteorological knowledge of storm types may be in the background but perhaps not explicitly modeled.

The most challenging research objectives concern the probing of data-generating processes with a view to enhanced understanding and to studying causality in one of the senses of that word; for a discussion of various views of statistical causality, see Cox and Wermuth [(1996), Section 8.7 and (2004)]. The fairly recent surge of interest in causality in statistical work has been very fruitful. The interventionist, counterfactual or potential outcomes formulation, emphasized in work stemming from Rubin (1974), is undoubtedly valuable in many fields of application, not least in clarifying which variables are to be regarded as potentially causal and in helping focus objectives. In the natural sciences the word causal tends to be used sparingly and, when it is used, means something like having an evidence-based interpretation in terms of an underlying process. Similar views have been suggested in sociology [Goldthorpe (2007)] and in demography [Ni Bhrolcháin and Dyson (2007)]. The latter authors, having developed a set of guidelines very similar to but rather more comprehensive than those of Bradford Hill (1965), widely quoted in epidemiology, suggest that three of the guidelines are essential. These are time ordering, evidence of mechanism and uniqueness of interpretation. Important as is the notion of causality, claims of having established it, in particular, from single observational studies, should be made with caution, in particular in those fields where there may be strong public policy or health implications.

Even in complex investigations, an ideal situation is that study objectives are specified by a series of focused research questions (or sometimes hypotheses). Often a focused question can initially or eventually be formulated by a one-dimensional parameter, implying that in statistical theory some emphasis should be placed on one-dimensional parameters of interest. In an older statistical tradition this was represented by the goal of breaking sums of squares of main interest in an analysis of variance table into single orthogonal degrees of freedom. The orthogonality now seems a rather secondary aspect but the one-dimensionality, if achievable, is important.

Even though the broad qualitative objectives of an investigation may be clear, nevertheless precise formulation of research questions may require considerable discussion. Interplay between subject-matter and statistical considerations may be crucial. Indeed, in some contexts it may be at this stage that statistical arguments are of most importance.

**3. Study design.** Next consider design. It is a truism that the design of a study, of whatever kind, is crucial. A seriously defective design may be incapable of rescue even by the most ingenious of analyses; a good design may lead to conclusions so clear that simple analyses are enough. Design issues range from the specific choice of material where the investigation involves secondary analysis of already collected data, through the various kinds of observational study to experiments, especially randomized experiments, where the investigator may have close to total control over the system under investigation. The possibility of mixtures of various types of study is important.

Objectives of design are essentially the same whatever the type of study, namely, to address the research questions so far as feasible free of bias, and with modest random error of estimation, as usually best assessed by a standard error, a width of confidence interval or posterior distribution. Note that initial assessment of precision is simpler by specifying a target standard error than by an essentially equivalent specification of power. The standard error emphasizes estimation rather than significance testing and power, unlike standard error, is of little direct relevance for analysis.

These aims are to be achieved with reasonable economy of resources. The general principles for doing all this are firmly established. The central challenge in virtually all instances is to adapt these principles to take account of practical constraints that are inevitably present. Simplicity is important. In particular, it is not helpful to identify statistical design of experiments with the use of combinatorially complex structures.

An important issue in design of all kinds involving relatively complicated situations not easily or quickly replicable is to consider the various patterns of outcome that may arise and to check that information will be available to interpret each such pattern if realized. This is not the same as assigning each such pattern a prior probability, although some may find it helpful to do that. Rather, it should include patterns that are thought unlikely, for example, effects in the opposite direction from those very confidently predicted. Experience suggests that, nevertheless, quite often patterns occur that had been totally overlooked initially, patterns possibly of great interest if real. How to deal with this possibility is taken up again in the section on analysis.

The aim in design is to achieve a secure investigation in which the analysis and its interpretation require as few external assumptions as possible. This contrasts with but does not conflict with R. A. Fisher's aphorism [Cochran (1965)]: *make your theories elaborate*. That is, assemble evidence of different kinds bearing on the topic under study.

Teaching design to students of statistics in a unified way is difficult; moreover, teaching design of experiments can all too easily drift into accounts of that traditional art-form, the analysis of variance table, a closely related but far from identical topic. Such tables, considered as capturing structure of degrees of freedom, are, however, crucial to the understanding of many more complicated designs.

Much the most interesting aspect of the design of experiments is actually helping design an experiment! It might be helpful if the new journal had from time to time accounts of the design phase of an investigation written in a form useful for students and giving some detail of the kind of interplay that goes on between subject-matter and statistical considerations, as well as of any novel features arising. For some of these issues, see Robinson (2000).

**4. Measurement.** Particularly in fields in which there is not a long tradition of measurement, careful study of measurement processes and protocols will be important and often crucial to obtaining solid conclusions. What do the data that one ends up analyzing actually mean?

The following criteria may be proposed for measurements under the headings below:

- Relevance.

- Precision.
- Effectiveness of implementation.
- Side effects of measurement.

For relevance, the measurements should capture the essence of the issues under study, be formed from a reasonable but not excessive number of distinct features and, particularly in a context with a solid foundation, allow connection with the subject-matter base of the field, for example, the laws of classical physics or quantum theory. As regards the number of component features, it is absurd, for example, to suppose that the economic health of a nation or the standing of a University or the health-related quality of life of a patient can for all purposes be summarized in a single number. In some contexts the primary measurements such as spectra or functional responses are themselves complex and require reduction in a preliminary phase of analysis. The broader issues of interpretability correspond to what psychometricians call the need for face and concept validity.

For precision, the requirement is that under stable conditions repeat measurements by the same or different observers give essentially similar answers. There is, of course, a very extensive statistical literature on how such issues should be addressed, estimating relevant components of variance, and on related matters, concerned, for example, with the recalibration of instruments. In the physical sciences metrology has brought the standardization of many key measurements to a very high level of precision. This high precision does, however, not eliminate the need to take account of complex and largely random patterns of real variability that may exist. It is noteworthy that in some areas of medicine, for example, there has not been a comparable effort on measurement issues.

The third requirement is that of effectiveness and economy of implementation. There is in some fields a very serious clash at the design phase between the need to collect wide-ranging data in order to be prepared for potential difficulties of interpretation and the expense and loss of quality that may result from aiming to collect too much data. That data quality may drop if too much is attempted is widely recognized. Recording some information only on appropriately chosen subsamples may be part of the answer.

The final requirement, less widely relevant than the others, is concerned with limiting the impact of what in the context of modern physics might be called the Heisenberg effect, that is, with the effect on the system under study of the measurement process itself. Thus, observers of animal behavior commonly recognize the need for considerable care to avoid distorting the phenomenon under study.

Stevens (1951) introduced a widely-referred to typology of measurement scales into ratio, interval, ordinal, etc. This was criticized in a social science context in an influential book by Duncan (1975), in particular, for undervaluing simple counting. Much statistical practice, correctly in my view, treats Stevens' classification as too prescriptive. Thus, data on the Mohs ordinal scale of hardness might very well be analyzed as an interval scale, for example, by linear least-squares regression. The spacings of the scale are not arbitrary as would be the implication of treating the scale formally as wholly ordinal. If reassurance is required, tests for the appropriateness of the scaling can be applied [Cox and Wermuth (1994)].

More important than the classification by measurement scales is often a classification by purpose. A simple division fairly widely applicable is into the following:

- outcome or response variables,
- intermediate variables,
- explanatory variables, themselves classified into the following:
  – primary explanatory variables (treatments or risk factors) really or conceptually capable of manipulation and, hence, in a relevant sense, potentially causal,
  – intrinsic (or background or context) variables,
  – nonspecific variables such as blocks in a randomized block experiment, centers or labs in an multi-center study.

Particularly in relatively complex observational studies, clarity in interpretation and design is aided by a graphical representation in which each variable is represented by a node of a graph. This is discussed further below.

Another important property is that some kinds of measurement are physically additive in a meaningful sense. In the physical sciences they are called extensive. The additivity implies that arithmetic means on the original scale have special virtue whatever the distributional shape. Thus, quantity of heat is extensive, temperature is not; the temperature of two bodies allowed to come to thermal equilibrium with one another is not in general the mean or the sum of their individual temperatures. That counting heads is extensive whereas log rates of occurrence are not describes the contrast in epidemiology between assessments by absolute and by relative risks. More broadly, in fields where primary measurements are related to the units of mass, length and time, dimensional analysis of fitted models is important and units should be chosen, usually in accord with the SI metric conventions, to produce, for example, regression coefficients of reasonable numerical magnitude. Dimensionless parameters are particularly appealing. Relative stability over different populations, an ultimately empirical issue, in principle, should indicate the preferred description, for example, of risk in epidemiology.

Yet other ways of describing measurements are as primary, or what are sometimes called pointer readings, or as derived from combinations of primary measurements. Sometimes these derived variables are determined from prior considerations or by tradition, as in the use of body mass index, that is, weight divided by height squared, in studies of obesity. In other cases an important task of preliminary analysis may be to determine appropriate derived variables. There are also important issues connected with the use of surrogate variables, especially surrogate response variables and, of course, also of representing hidden or latent variables.

For a wide-ranging account of measurement, see Hand (2004).

Issues of data storage and management are defined to be outside the scope of the present discussion. While many studies nowadays involve very large amounts of data, this does not in general imply large amounts of information about the research questions of interest. Clearly, new issues arise from the prevalence of large data matrices and the desirability of graphical analysis of multidimensional data. Nevertheless, it is tacit in the following discussion, and certainly open to dispute, that the broad approaches desirable to analysis and interpretation have not been radically changed by the capability to handle very large amounts of data, however, much approaches to implementation have been and are being revolutionized.

## 5. Analysis.

5.1. *Phases of analysis.* Analysis of data is divided broadly into three or sometimes four phases. The three phases are data editing and quality control, preliminary somewhat informal analysis and more detailed analysis. Sometimes, however, there may be a critical phase of preliminary analysis and data processing in which complex primary data are converted into a form suitable for direct analysis and interpretation. The present discussion concentrates on the last phase, namely, that of more detailed analysis. While much of the specific discussion is set out for analysis based on probabilistic models, many of the more important points apply also to analyses based on algorithmic approaches such as the various forms of cluster analysis, correspondence analysis and multidimensional scaling and to situations where empirical prediction is a central objective.

It is scarcely possible in a short paper like this to do more than touch on one or two of the more general issues concerning analysis.

5.2. *Transparency.* An important aspect of analysis, difficult to achieve with complex methods, is transparency. That is, in principle, the pathway between the data and the conclusions should be as clear as is feasible. This is partly for the self- education of the analyst and also is for protection against errors of data transcription, undue dependence on suspicious values and so on. It is important also for presenting conclusions.

Checks of the more important results of statistical analysis are important, if at all possible by using a detailed numerical method different from that employed initially. While nowadays the possibility of unsignalled numerical-analytic error is usually relatively small, other sources of error certainly remain. Transparency is partly addressed to the same issue as checking, namely, of achieving security by an independent route. Thus, results from numerical procedures may be related to the data graphically. Transparency strongly encourages the use of the simplest methods that will be adequate.

5.3. *Fragmentation.* Transparency may also be aided by dividing the data into meaningful sections, initially for separate analysis. This is particularly appropriate in two rather different circumstances. In one there is a preliminary suggestion of a considerable number of interactions involving an intrinsic variable such as gender. Then a stage of analysis in which men and women are studied separately may be wise. Another possibility is that the data arise from different time periods or different sites. There is then a choice between, on the one hand, analysis including interaction terms with, say, times and, on the other hand, analyzing the time periods in the first place separately.

Yet fragmentation of the final synthesis is to be avoided if possible. For example, given a number of different but similar time series to be fitted by autoregressive-moving average models, it might be reasonable to fit them first separately choosing the order of representation by one or other of the methods available. It would then usually be helpful to choose one type of model, possibly one of degree high enough to fit all the series adequately, and then to fit the same type of model to all, thus enhancing comparability and synthesis of conclusions. Again, some computer packages implicitly encourage the production of large numbers of $p$-values testing highly related hypotheses on similar but separate sets of data. This is usually not only irrelevant but misleading in that it encourages comparison between the "significant" sets and the "insignificant" sets. The aim should be to encapsulate the issue under study in a single parameter capturing an effect constant across sets and, if that is not possible, toward explaining differences in effect between sets.

5.4. *Model definition and choice.* Algorithmic methods of analysis and purely empirical forecasting procedures are not based on an explicit probability model, although it may sometimes be enlightening to consider qualitatively what kind of model might have led to the procedure involved. The corresponding question for analyses based on a formal model concerns the direct descriptive reasonableness of the procedure involved, regardless of the specific formal model employed.

The very powerful idea of analyses explicitly based on a probability model covers most of the work published in the statistical literature. The following discussion is largely in terms of parametric models.

The primary role of such models is to represent, of course in idealized form, the occurrence in the real world of both systematic and unexplained variability. Probability is best interpreted as a limiting frequency under repetition which may be approximately realizable or may be hypothetical or even highly hypothetical, as with applications to macro-economics or to literary studies. The limiting process is, in principle, no different from that in many areas of applied mathematics, for example, in the definition of fluid density in continuum mechanics. The idea is that the model should capture the essence of the data-generating process decoupled from the accidental aspects of the specific data under analysis. It is an approximate description of something objective.

Models may be purely descriptive representations of commonly occurring patterns of systematic and haphazard variability. Alternatively, they may contain substantial elements of subject-matter theory or experience [Lehmann (1990), Cox (1990)]. Stemming from this distinction, there are two differing trends in model construction. One is toward ever more general families of model which, with appropriate software, allow empirical fitting of complex systems in many fields. The other trend is toward developing more and more specific models taking into account background knowledge appropriate to individual situations.

The need for formal theory in statistics comes from a number of directions. One is the need to provide a basis for assessments of uncertainty. These need to be as objective as is possible and provide both warning against effects that may arise purely by the play of chance and also security to justified conclusions. A second is the need for a systematic approach to the analysis of new models and for some synthesis of the large body of special techniques of analysis now available.

The choice of parameters, especially the parameters of interest, is critical. For interpretative purposes, the key issues are that parameters of interest should be such that, separately for each research question of concern:

- there is a parameter addressing that question,
- each such parameter has a clear subject-matter interpretation, such as being a rate of change or a contrast between levels of a qualitative variable,
- interpretation is largely retained under perturbation of the original model.

For nuisance parameters such issues as reliable convergence of methods of estimation become relatively more central.

5.5. *Graphical models of dependency.* Substantive models often stem from differential equations that represent the development of a physical or biological system. These equations lead either to a stochastic model that can be fitted to data or, more commonly, to a deterministic model which can be supplemented by additional error terms. In some of the fields in which such development is not available an approach stemming from the path analysis of Sewall Wright (1921) may serve a similar purpose. Graphical Markov models aim to represent a data-generating process by a graph in which each node represents a variable and dependencies of various kinds between pairs of variables are represented by edges, some directed some undirected, so that a missing edge represents a conditional independency. Such graphs have been studied from several viewpoints, some computational; for an introductory account focusing on statistical analysis, see Cox and Wermuth (1996).

Advantages of this approach include a graphical representation of possibly relatively complicated dependencies as an aid to interpretation, and the ability to deduce further independencies implied by a particular model. Recent developments allow the deduction of properties of related systems in which some variables are missing or their status as explanatory and response variables changed.

5.6. *Hierarchical issues.* Haphazard variation is often structured, for example, variation in individuals close in space or time may well be correlated. There are many fields also where variation is hierarchical and recognition that haphazard variation may enter at several levels is often crucial. In an educational study, interest may lie in school areas, in schools, in classes within schools and in individual pupils. In a community randomized trial there are likely to be data on communities, families within communities and individuals within families. There may now be a tension. Attempts at causal interpretation in the sense described above demand analysis, and probably modeling, at more than one hierarchical level. On the other hand, in a primary analysis, say, of the effect of an intervention at community level, the unit of study is the unit of randomization, the community. Analyses at a sub-community level have the status of an observational study, suggesting that modeling at the sub-community level may in the first instance be unnecessary. This leads to a so-called principle of minimal modeling, namely, to confine the probabilistic modeling to those features directly necessary for interpretation.

In more general studies of dependence the recognition that dependencies at different hierarchical levels may be different may be crucial. There is an extensive literature, stemming in a sense from an older literature on analysis of variance. Appreciation is important of when a complex structure is and when it is not germane to possible biases in estimation and also to underestimation of random errors of estimation.

5.7. *Empirical model choice.* There is an extensive and ever-growing literature on empirical model selection, that is, choice of a particular model from within some specified family. In its simplest manifestation this is the family of linear regressions on a specified and possibly large set of explanatory variables. Much of this literature centers on use of significance or criteria such as AIC or BIC, combined with a forward or backward procedure, to choose a single relatively simple model.

Those discussions are mostly focused on prediction. For that purpose a limitation of many analyses is that the stability of the predicting equation over changes likely to occur in repeated application is not explicitly considered.

For interpretative purposes, there are three important additional considerations. First, if there are different essentially equally well-fitting representations of the data, it will be desirable to know and usually to report this. Second, account should be taken of the differences between the various types of explanatory variables. Which primary explanatory variables are included, if a choice has to be made, is usually crucial. For example it would be absurd to exclude a variable representing a key objective of the investigation because its effect was insignificant. The selection of background variables is less critical provided major biases are avoided and the central issue is the stability of the effects of primary variables. Third, because, particularly in complex systems, total linearity is implausible, tests for nonlinearity and interaction are desirable at some stage of analysis.

There may be an apparent conflict with the widely-held belief in parsimony. The object is to answer research questions, ideally well focused in the simplest form that captures their essence, allowing for reasonable protection against bias and for precision enhancement. Yet that is not the same as choosing the model involving the smallest number of adjustable parameters. Reality can be complicated and one wants assurance that any simple answer achieved both addresses the research question at issue and provides as secure an answer as possible.

5.8. *Local and external precision assessment.* Clearly, a very distinctive feature of statistical inference lies not only in its ability to summarize data in what should be an enlightening way, but also in the ability to assess the uncertainty of resulting conclusions. Indeed, some would regard the assessment of uncertainty as the defining feature of statistical inference. It would be out of place here to discuss the various ways in which that uncertainty may be described.

Assessment of precision, leading therefore to assessment of uncertainty, can be regarded as coming from two sources, not always clearly distinguishable. One is internal or local. In fitting a model to a single relatively homogeneous set of data estimates of precision typically come from an estimate of variability derived from variation of the observations around fitted values, for example, from a residual mean square in a least squares regression, or directly from repeat observations on the same individual, or from an assumption of binomial or Poisson variability that relates variance to mean. Such estimates are dependent on relatively strong model assumptions, especially of independence.

Such estimated precisions may be contrasted with those that come essentially from estimating the parameter in question on largely or wholly separate sets of data. For example, there may be data from different centers or laboratories for each of which the parameter can be estimated separately. The variability of the overall estimated effect, assessed formally by an interaction term of the broad form treatments × centers, leads to a more direct estimate of stability. Realistic differences between centers enhance the generalizability of the conclusions, a standard principle of classical experimental design, emphasized also in Taguchi's system of industrial design. The randomized block design, and, in particular, the matched pair design, forms its simplest illustration.

5.9. *Reformulation of objectives.* The design of any potentially important study includes not only a broad specification of the intended analysis but also a check of the ability to interpret the various possible patterns that may emerge from the formal analysis. It will virtually always be desirable to set out a plan for analysis in written form. This is much less, however, than requiring specification in detail of the analysis that must be done regardless of the data, although that may sometimes be necessary for administrative, regulatory or quasi-political reasons. There are two broad ways in which an initial analysis may be adapted in the light of the data. One involves no change in the research questions under study but a technical change in the specification of the probability model. Such changes may involve a change in the definition of the parameter of interest, following, for example, nonlinear transformation of a response variable or a modification of the assumed form of a dose- or stimulus-response curve. A quite different issue is that unexpected features of the data may suggest a major change of objective.

So long as the first type of change is based on improving model fit, not on maximizing apparent statistical significance, no change in formal assessment of uncertainty appears needed. The adaptation can be regarded as the result of an informal maximum likelihood analysis over an essentially orthogonal parameter representing model type. Without some such argument, a great many analyses would lose their formal justification.

If, however, there is a radical change in direction, the situation is quite different. Standard procedure in those fields where investigation can be repeated fairly quickly is to set up a confirmatory study. In other fields, while repetition may be impracticable, it may be possible to obtain separate information that is some check on the new development. If the new idea arose out of systematic search through a set of possibilities that can be approximately specified, then corrections of Bonferroni type (or approximate specification of prior probabilities) may be possible. But for strikingly new developments, this is not feasible. What was the set of possibilities over which search took place? The prior probability as assessed beforehand of a totally unanticipated effect is essentially zero. The prior probability assessed in the light of the data may be high, but this provides little or no reassurance against following false trails! The position seems to be that formal statistical assessment of security from the originating study alone is impossible; reassurance or refutation must come from combination with other sources of information.

5.10. *Issues concerning small probabilities.* In the majority of assessments of uncertainty once probabilities are much below $10^{-2}$ or $10^{-3}$, say, the precise value is likely to be quite unreliable. This applies to posterior probabilities or to derived values such as levels of significance. Even if the assumptions underlying the values are well founded, they cannot be pushed too far and sources of uncertainty not accounted for in the calculation may become predominant. There are, however, a number of circumstances under which very small probabilities, possibly $p$-values, are important.

These situations seem to be of two broad types. Some are direct calculations about possible outcomes, for example, calculations of the reliability of complex systems or estimation concerned with extreme daily rainfalls for a specified reservoir catchment area. Here it is required to consider events with probabilities of the order of $10^{-4}$ per year, that is, roughly $1/3 \times 10^{-6}$ per day. The best that can be done here is to use data collected over many sites and as long a time period as is available to check that methods give reasonable results for extreme levels as near as possible to those required, aiming to reduce the element of extrapolation.

A second type of problem concerns assessment of uncertainty attached to the outcome of extended searches. The detection of candidate genes for rare diseases is one application. Another is in particle physics where the search for the Higgs boson involves looking for an occurrence against an enormous uninteresting background. For an account of the wide ranging analytical issues involved, see Cranmer (2006). There may be an approach to a consensus in that field that five sigma significance is required to support claims. This is presumably a rather arbitrary combination of a safety factor for unassessed sources of error plus an allowance for selection. The formal Gaussian tail probability is about $10^{-7}$. The situation in genetics may be less critical, in particular, because of the possibility of replication. This means that the process may be better represented as multi-stage selection in which emphasis is placed not on intermediate $p$-values but on the soundness of the ultimate outcome.

**6. Interpretation.** The borderline between analysis and interpretation is not clear-cut but corresponds essentially to that between the Results and Discussion sections of the conventionally laid out paper in the natural sciences. One theme of the previous section is implicitly the narrowing of the gap between the two phases. At a fairly simple level, it is important to present the conclusions in a form in which relatively direct interpretation is encouraged. Thus, logistic regression is a widely used technique for the analysis of binary outcomes and the conclusions are commonly presented in a table of regression coefficients with standard errors. These are often quite unenlightening in terms of interpretation. For example, it may be hard from the regression coefficients to assess the effect on the probability of success of an individual of making realistic changes in key explanatory variables. Clearly, the more the statistical model used as a basis for a formal analysis involves subject-matter ideas, the easier the transition to interpretation will be.

There is a tension, although not a clash, between two statistical themes. One is the notion of the self-contained investigation. The most obvious example is the randomized factorial experiment in which, by a judicious mixture of randomization, replication, error control and the factorial principle, multiple research questions can be addressed with security and without major external uncheckable assumptions. Yet R. A. Fisher's aphorism mentioned in Section 3 is closer to the spirit of scientific investigation in which security comes not only from individual pieces of investigation but also from consolidating information of very different kinds into a consistent whole.

In the final discussion, whether it is verbal or in a scientific paper, it is desirable both that the nature of the conclusions and of uncertainties attached to them are clear and also that the route to the conclusions is at least broadly understandable. The issue of transparency discussed in Section 5.2 is very relevant.

For studies with a directly decision-making focus, the issues analogous to interpretation are those of implementation and, preferably, of monitoring of the outcome of the implementation.

**7. Conclusion.** This essay has been based on the ordering:

- formulation of objectives,
- design,
- measurement,
- analysis of data,
- interpretation.

This ordering, while natural enough, may easily become distorted. From the perspective of an individual statistician, entry may be at the analysis phase. It may then be quite a challenging task to uncover what aspects of the conceptually earlier steps of design and measurement are critical to analysis, as well, crucially, as clarifying the research questions of concern. A different reordering, the important topic of the reformulation of research questions in the light of analysis, has been discussed above. Quite commonly, interim analysis may suggest sharpening the formulation of the research question. Indeed, the essentially iterative character of analysis and interpretation both within and between studies is crucial [Box (1976)]. The time scale on which the iteration takes place clearly varies greatly between different fields.

It may be felt that this account has overemphasized the "soft," that is, non- mathematical, side of applied statistics at the expense of the "hard," that is, mathematical, challenges of methodological development. To think that "soft" in this context means "easy" would be a dreadful mistake.

In many fields of application a quite crucial role is played by standard software packages. Methodological development is, however, often triggered by the recognition of special features in specific applications that may show that the "standard" analyses are either inapplicable or not quite appropriate. It is this recognition, preferably of broad families of problems where new formulations are needed, that gives the purely methodological side of the subject its intellectual vigor, fascination and importance.

A final general point is that this discussion has been about statistics not about statisticians. How individuals work most fruitfully in what, except possibly for theoretical development, is typically a largely collective enterprise raises many further issues.

That the new journal will set high standards is beyond doubt. It is almost inevitable that detailed methods of analysis and examples thereof, hopefully with discussion of the interplay between statistical analysis and subject-matter interpretation, will form a dominant theme, but I hope that the broader issues touched on in this essay are not ignored.

**Acknowledgments.** I am very grateful to Brad Efron for inviting these remarks and to Sarah Darby, David Firth, Rolf Sundberg and Nanny Wermuth for very constructive comments on an earlier version.

Nuffield College
Oxford OX1 1NF
United Kingdom
E-mail: david.cox@nuffield.ox.ac.uk